\documentclass[aps,amsmath,amssymb, prc,twocolumn,superscriptaddress]{revtex4-1}

\usepackage{graphicx}
\usepackage{dcolumn}
\usepackage{bm}
\usepackage[utf8]{inputenc}
\usepackage[ngerman,english]{babel}
\usepackage{float}
\usepackage{multirow}

\begin{document}


\title{Possible experimental signature of octupole correlations in the 0$^+_2$ states of the actinides}


\author{M. Spieker}
\email[]{spieker@ikp.uni-koeln.de}
\homepage[Member of the Bonn-Cologne Graduate School of Physics and Astronomy]{}
\affiliation{Institut f\"{u}r Kernphysik, Universit\"{a}t zu K\"{o}ln, Z\"{u}lpicher Straße 77, D-50937 K\"{o}ln, Germany}

\author{D. Bucurescu}
\affiliation{Horia Hulubei National Institute of Physics and Nuclear Engineering, Bucharest, Romania}

\author{J. Endres}
\affiliation{Institut f\"{u}r Kernphysik, Universit\"{a}t zu K\"{o}ln, Z\"{u}lpicher Straße 77, D-50937 K\"{o}ln, Germany}

\author{T. Faestermann}
\affiliation{Physik Department, Technische Universität München, München, Germany}

\author{R. Hertenberger}
\affiliation{Fakultät für Physik, Ludwig-Maximilians-Universität München, München, Germany}

\author{S. Pascu}
\affiliation{Institut f\"{u}r Kernphysik, Universit\"{a}t zu K\"{o}ln, Z\"{u}lpicher Straße 77, D-50937 K\"{o}ln, Germany}
\affiliation{Horia Hulubei National Institute of Physics and Nuclear Engineering, Bucharest, Romania}

\author{S. Skalacki}
\affiliation{Institut f\"{u}r Kernphysik, Universit\"{a}t zu K\"{o}ln, Z\"{u}lpicher Straße 77, D-50937 K\"{o}ln, Germany}

\author{S. Weber}
\affiliation{Institut f\"{u}r Kernphysik, Universit\"{a}t zu K\"{o}ln, Z\"{u}lpicher Straße 77, D-50937 K\"{o}ln, Germany}

\author{H.-F. Wirth}
\affiliation{Fakultät für Physik, Ludwig-Maximilians-Universität München, München, Germany}

\author{N.-V. Zamfir}
\affiliation{Horia Hulubei National Institute of Physics and Nuclear Engineering, Bucharest, Romania}

\author{A. Zilges}
\affiliation{Institut f\"{u}r Kernphysik, Universit\"{a}t zu K\"{o}ln, Z\"{u}lpicher Straße 77, D-50937 K\"{o}ln, Germany}


\date{\today}

\begin{abstract}

$J^{\pi}$= 0$^+$ states have been investigated in the actinide nucleus ${}^{240}$Pu up to an excitation energy of 3~MeV with a high-resolution {\it (p,t)} experiment at $E_{p}$=~24~MeV. To test the recently proposed $J^{\pi}$=~0$^+_2$ double-octupole structure, the phenomenological approach of the {\it spdf}-interacting boson model has been chosen. In addition, the total 0$^+$ strength distribution and the $0^+$ strength fragmentation have been compared to the model predictions as well as to the previously studied {\it (p,t)} reactions in the actinides. The results suggest that the structure of the 0$^+_2$ states in the actinides might be more complex than the usually discussed pairing isomers. Instead, the octupole degree of freedom might contribute significantly. The signature of two close-lying 0$^+$ states below the \mbox{2-quasiparticle} energy is presented as a possible manifestation of strong octupole correlations in the structure of the 0$^+_2$ states in the actinides.

\end{abstract}


\maketitle


The nature of the first excited $K^{\pi}=~0^+,~J^{\pi}=~0^+$ (0$^+_2$) states in the even-even actinides has been controversially discussed for decades. Extensive experimental studies had shown an asymmetry between the population in {\it (p,t)} and {\it (t,p)} reactions~\cite{Maher72, Fried74, Cast72}. While large {\it (p,t)} cross sections were observed ($\sigma_{0^+_2}/\sigma_{\mathrm{gs}} \thickapprox 15\%$) \cite{Maher72, Fried74}, no significant {\it (t,p)} cross sections ($\sigma_{0^+_2}/\sigma_{\mathrm{gs}} < 2\%$) were seen in the studies of Casten {\it et al.}~\cite{Cast72}. Ragnarsson and Broglia proposed to interpret these states as pairing isomers having a smaller neutron pairing gap $\Delta_{n}$ than the ground state itself \cite{Ragn76, Rij72}. These isomers should be present in the case of an inhomogeneity of weakly coupled prolate and oblate levels around the Fermi surface for comparable monopole and quadrupole pairing strengths. The experimental signature of these pairing isomers would be large {\it (p,t)} cross sections and almost vanishing {\it (t,p)} cross sections~\cite{Ragn76}. Furthermore, Rij and Kahana predicted a negligible population of these states in single-neutron transfer reactions~\cite{Rij72}. However, in ${}^{240}$Pu a strong {\it (d,p)} population of the $0^+_2$ state ($\sigma_{0^+_2}/\sigma_{gs} \thickapprox 18\%$) was observed~\cite{Fried73}. In addition to the peculiar situation of an inhomogeneous distribution of oblate and prolate single-particle orbitals around the Fermi surface, the octupole degree of freedom contributes significantly to the low-energy spectrum of the actinide nuclei~\cite{Butl96, Shel2000}. The enhanced octupole collectivity, which has been observed in several experimental and theoretical studies, has been mainly attributed to the strong octupole interaction involving unique neutron and proton single-particle orbitals differing by $\Delta j=~\Delta l$~=~3. These are the $j_{15/2}$ and $g_{9/2}$ orbitals ($N~\sim$~134) for the neutrons, and the $i_{13/2}$ and $f_{7/2}$ orbitals ($Z~\sim$~88) in the case of the protons. One of the most striking signatures of enhanced octupole collectivity has been the observation of alternating-parity bands where negative-parity states are connected to positive-parity states via enhanced $E1$ transitions and vice versa. Nuclei which exhibit such an alternating-parity band already at low spins ($J \sim$ 5) have often been considered as the best candidates to exhibit static octupole deformation already in their ground state (see, e.g., ${}^{226}$Th~\cite{Butl96} and ${}^{224}$Ra~\cite{Gaff13}). 

\begin{figure*}[!t]
\centering
\includegraphics[width=0.765\linewidth]{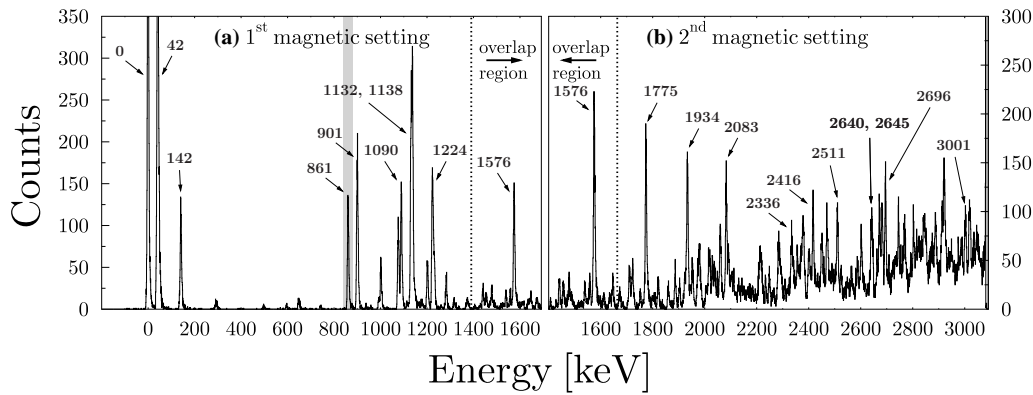}
\caption{\label{fig:spectrum} The ${}^{242}$Pu{\it (p,t)}${}^{240}$Pu spectra at 10$^{\circ}$ for both magnetic settings which were used for the energy calibration of the focal plane detection system. The overlap regions are marked by dashed lines. Prominent peaks are highlighted with their excitation energy in keV. In addition, the proposed double-octupole phonon $J^{\pi}$=~0$^+_2$ state~\cite{Wang09, Jolo13} is marked with a grey shaded area.}
\end{figure*}

In the even-even Pu isotopes the strongest octupole correlations have been found for ${}^{240}$Pu by means of an alternating-parity band at high spins ($J \sim 20$)~\cite{Wied99}. Using two-center octupole wave functions in the framework of supersymmetric quantum mechanics, Refs. \cite{Jolo11, Jolo12} explained the experimental data as a second-order phase transition from an octupole-nondeformed to an octupole-deformed shape. Recently, the data of a consecutive high-statistics ``unsafe'' Coulomb excitation experiment has been published~\cite{Wang09}, where the $K^{\pi}=$~0$^+_2$ rotational band was investigated up to highest spins ($J^{\pi}=~30^+$). Enhanced $E1$ transitions were observed which deexcited its high-spin members exclusively to the one-octupole phonon band ($K^{\pi}=$~0$^-_1$). Following the proposed concept of multi-phonon condensation of Ref. \cite{Frauend08}, the experimental observations in ${}^{240}$Pu \cite{Wied99, Wang09} were explained in terms of the condensation of rotation-aligned octupole phonons at high spin. As a consequence, Ref. \cite{Wang09} proposed the $K^{\pi}=$~0$^+_2$ rotational band as a candidate for the double-octupole band. Earlier, Chasman had noticed that the addition of particle-hole octupole-octupole forces to the conventional pairing forces might also explain the strong {\it (p,t)} population of 0$^+_2$ states in the actinides~\cite{Chas79}. He had found enhanced octupole correlations in the 0$^+_2$ states of some U isotopes. The recent experimental studies of monopole excitations in deformed nuclei~\cite{Lesh02, Wirt04, Meye06, Lesh07} have shown that with state-of-the-art experimental setups new structure information can be gained about many excited $J^{\pi}$=~0$^+$ states up to an excitation energy of about 3 MeV. It is important to note that also the {\it spdf}-interacting boson model (IBM) predicts a significant contribution of the double-octupole phonons to the structure of the excited $J^{\pi}$=~0$^+$ states in the actinides~\cite{Zamf01, Zamf03, Wirt04, Levon09, Levon13}. The new interpretation of the 0$^+_2$ state and its proposed structure in ${}^{240}$Pu~\cite{Wang09, Jolo13} therefore emphasizes the importance of the present experimental {\it (p,t)} study to further clarify the nature of the 0$^+_2$ states in the actinides and to further test the existence of low-lying two-phonon states in deformed nuclei~\cite{Solov1981}.

\begin{figure}[ht]
\centering
\includegraphics[width=.854\linewidth]{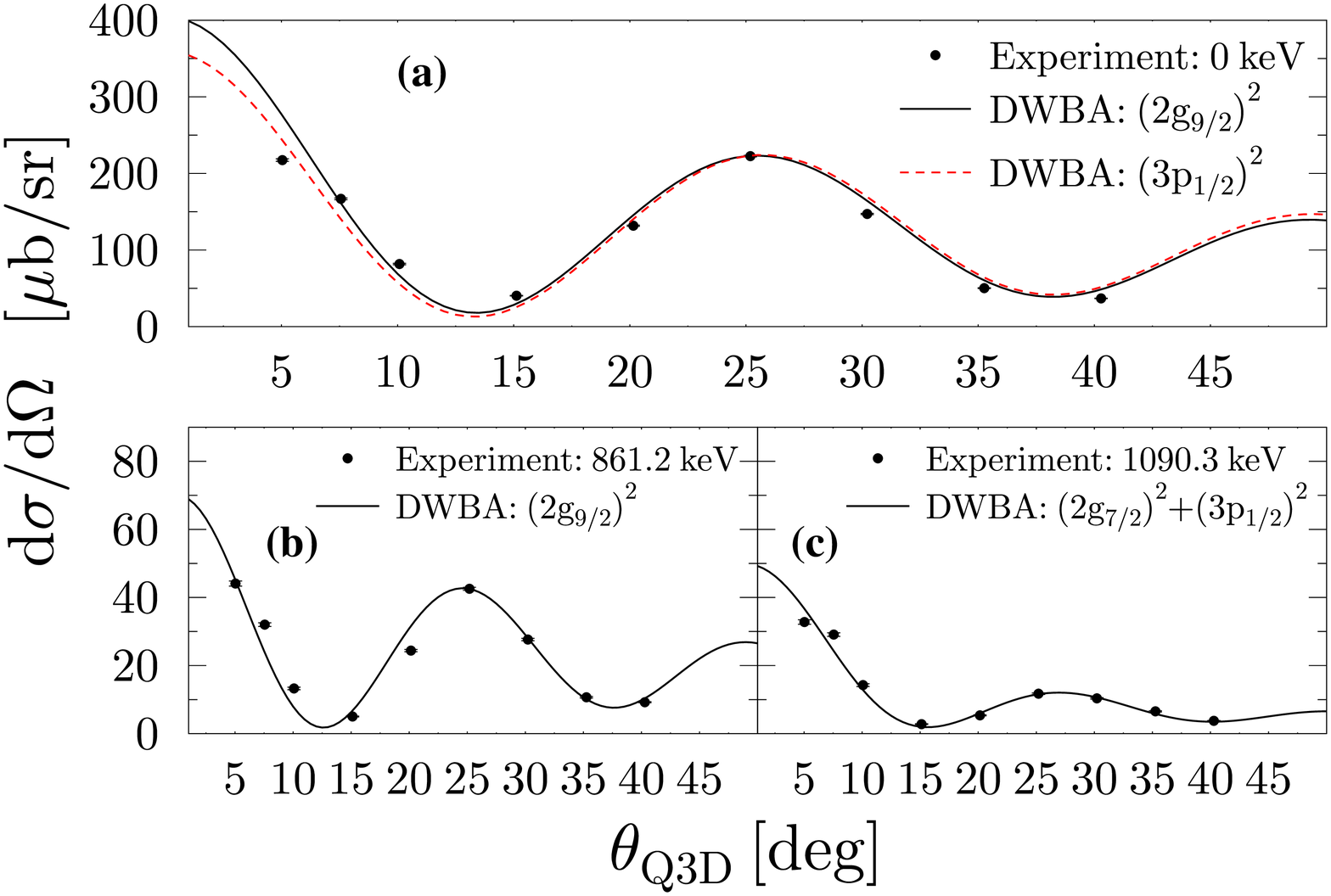}
\caption{\label{fig:gsangdist}(Color online) Angular distributions of the {\bf (a)} $J^{\pi}$=~0$^+_1$ ground state, {\bf (b)} $J^{\pi}$=~0$^+_2$ state at 861.2(1) keV , and {\bf (c)} $J^{\pi}$=~0$^+_3$ state at 1090.3(1) keV in ${}^{240}$Pu. The angular distributions for the corresponding $L$~=~0 transfer calculated with the CHUCK3 code \cite{CHUCK} are also shown. For the latter, two-neutron transfer configurations of orbitals close to the Fermi surface were chosen. For further details of the CHUCK3 calculations see, e.g., Ref.~\cite{Levon09}.}
\end{figure}

To study 0$^+$ states and other low-spin excitations in ${}^{240}$Pu a high-resolution {\it (p,t)} study was performed at the Q3D magnetic spectrograph of the Maier-Leibnitz Laboratory (MLL) in Munich~\cite{loef73}. A 120~$\mathrm{\mu g/cm^2}$ thick and highly-enriched ${}^{242}$Pu target (99.93$\%$, T$_{1/2}$=~3.75~$\times$~10$^5$ years) was provided by Oak Ridge National Laboratory, and was evaporated onto a 25~$\mathrm{\mu g/cm^2}$ carbon backing. The $E_{p}$=~24~MeV proton beam impinged onto the ${}^{242}$Pu target with an average beam current of 1~$\mu$A. The ejected tritons were bent to the focal plane detection system of the Q3D magnetic spectrograph, where they were unambiguously selected via {\it dE/E} particle identification \cite{Wirt00}. The energy calibration of the detection system was done in terms of well-known {\it (p,t)} reactions as presented in, e.g., Ref.~\cite{Levon09}. Figure~\ref{fig:spectrum} shows the excitation spectrum of ${}^{240}$Pu for the two magnetic settings at 10$^{\circ}$, which have been used to cover excitation energies up to 3~MeV. Former studies in the rare earths as well as in the actinides have shown that high-resolution two-neutron transfer experiments with an incident proton energy of $E_{p} \thickapprox$ 25~MeV and the use of small detection angles are excellent tools to study $J^{\pi}$=~0$^+$ states selectively (see, e.g., Refs.~\cite{Meye06, Wirt04}). As seen in Fig.~\ref{fig:gsangdist}, the $L$~=~0 transfers can be easily distinguished from other low-spin excitations by their steeply rising cross sections at small detection angles at these proton energies. To unambiguously assign spin and parity to $J^{\pi}$=~0$^+$ states, angular distributions were measured at nine laboratory angles ranging from 5$^{\circ}$ to 40$^{\circ}$ and compared to the distributions calculated by the CHUCK3 code \citep{CHUCK}. Except for the measurements at 5$^{\circ}$ (9.3 msr), the maximum Q3D solid angle of 13.9 msr was chosen. This procedure has already been successfully applied to the ${}^{232}$Th{\it (p,t)}${}^{230}$Th reaction~\cite{Levon09}. In total, 209 states have been identified in the present {\it (p,t)} study in ${}^{240}$Pu. Many previously known low-spin states have also been observed and their spin-parity assignments were confirmed. However, most of the populated states have been seen for the first time~\cite{Spiek13}. This manuscript concentrates on the experimental distribution of populated $J^{\pi}$=~0$^+$ states up to 3 MeV, and especially on the pattern of low- and close-lying 0$^+_2$ and 0$^+_3$ states in ${}^{240}$Pu. The occurence of this signature is systematically studied along the actinides, and a possible interpretation will be presented in the framework of the {\it spdf}-IBM.

The ground-state angular distribution of ${}^{240}$Pu serves as a benchmark, since the unique signatures of an $L$~=~0 transfer in the {\it (p,t)} reaction are clearly identified [see Fig. \ref{fig:gsangdist} {\bf (a)}]. In total, 17 states were firmly assigned $J^{\pi}$=~0$^+$ by the present study. Furthermore, four additional states were tentatively assigned $J^{\pi}$=~0$^+$. The first excited $J^{\pi}$=~0$^+_2$ ($K^{\pi}$=~0$^+_2$) state, which is the candidate for the double-octupole 0$^+$ state~\cite{Wang09}, is identified at an excitation energy of 861.2(1)~keV, while the second excited J$^{\pi}$=~0$^+_3$ state is observed at an excitation energy of 1090.3(1) keV in the present experiment. The latter was proposed to be a candidate for the $\beta$-vibrational bandhead~\cite{Hoog96}. Both spin-parity assignments $J^{\pi}$=~0$^+$ \cite{Sing08} are confirmed [see Figs.~\ref{fig:gsangdist} {\bf (b)} and \ref{fig:gsangdist} {\bf (c)}].

\begin{figure}[ht]
\centering
\includegraphics[width=0.82\linewidth]{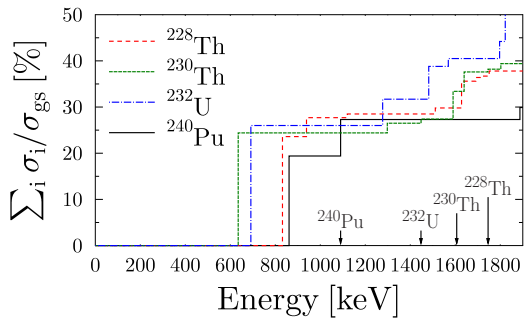}
\caption{\label{fig:runningsum}(Color online) Running relative transfer strength as a function of energy up to the 2QP energy (solid arrows) in ${}^{228}$Th. The neutron-pairing energy $\Delta_{\mathrm{n}}$ was calculated from the odd-even mass differences~\cite{Audi03}. Except for ${}^{240}$Pu, the data are from Refs.~\cite{Wirt04, Levon09, Levon13}.}
\end{figure}

The experimental integrated {\it (p,t)} cross sections are 173.75(7), 33.69(3), and 13.83(2)~$\mathrm{\mu b}$ for the ground state, 0$^+_2$ state, and 0$^+_3$ state, respectively. This results in relative {\it (p,t)} transfer strengths of $\sigma_{\mathrm{0^+_2}}/\sigma_{\mathrm{gs}}$~=~19.39(2)$\%$ and $\sigma_{\mathrm{0^+_3}}/\sigma_{\mathrm{gs}}$~=~7.96(1)$\%$. Only statistical errors are given. Systematic errors in the determination of the cross sections arise from the current integration in the Faraday cup and are roughly 10$\%$. In ${}^{240}$Pu, the total relative transfer strength adds up to 68.45(8)$\%$ [64.57(8)$\%$ for firm 0$^+$ assignments], and hence fits the systematics of Refs.~\cite{Wirt04, Levon09, Levon13}. As observed earlier~\cite{Maher72}, most of the strength is carried by the uniformly strongly excited 0$^+_2$ state. A striking observation of the systematic study is the pattern of the close-lying 0$^+_2$ and 0$^+_3$ states in ${}^{228}$Th ($\Delta E \thickapprox$~107~keV)~\cite{Levon13} and ${}^{240}$Pu ($\Delta E \thickapprox$~230~keV), while in ${}^{230}$Th and ${}^{232}$U these states are separated by $\Delta E \thickapprox$~600~keV~\cite{Wirt04, Levon09}. As seen in Fig.~\ref{fig:runningsum}, both 0$^+$ states lie below the 2-quasiparticle (2QP) energy and might therefore be of collective nature. Note that around the 2QP energy, excited 0$^+$ states are observed with a relative strength between 6$\%$ and 8$\%$. In ${}^{240}$Pu where the 2QP energy is the lowest ($2\Delta_{n} \thickapprox$~1090~keV), no additional strength is observed up to roughly 1.9~MeV~\cite{Spiek13}. As such, ${}^{240}$Pu marks an ideal case for studying the structure of close-lying 0$^+_2$ and 0$^+_3$ states.

\begin{figure}[ht]
\centering
\includegraphics[width=0.71\linewidth]{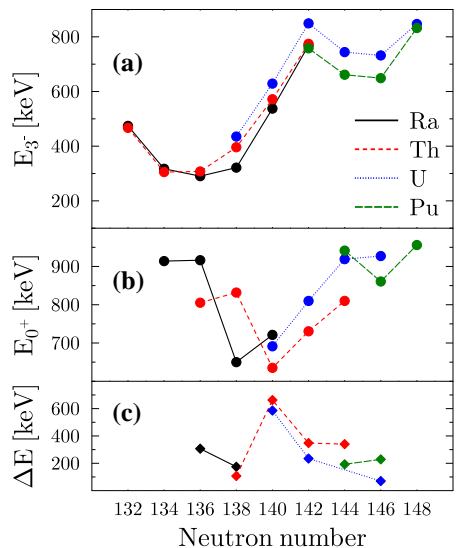}
\caption{\label{fig:actinidessystematics}(Color online) Evolution of the excitation energies of the {\bf (a)} $3^-_1$ and {\bf (b)} 0$^+_2$ states in actinides. {\bf (c)} Energy difference $\Delta E$ of 0$^+_2$ and 0$^+_3$ states where known~\cite{Wirt04, Levon09, Levon13, ENSDF}.}
\end{figure}

In Fig.~\ref{fig:actinidessystematics} $\Delta E$ has been systematically studied to test whether it could be used to gain further information about those states in the actinides, and to search for an indication of the double-octupole phonon structures at low excitation energies. In the Th and U isotopes, the maximum of $\Delta E$ seems to be correlated with the minimum energy of the 0$^+_2$ state [Figs.~\ref{fig:actinidessystematics} {\bf (b)} and \ref{fig:actinidessystematics} {\bf (c)}). Moreover, the minimum of $\Delta E$ appears to coincide with the minimum energy of the $3^-_1$ state [Figs.~\ref{fig:actinidessystematics} {\bf (a)} and \ref{fig:actinidessystematics} {\bf (c)}), which is an empirical measure for enhanced octupole collectivity. In the Pu isotopes, the minimum energy of the $3^-_1$ state might be more shallow~\cite{Shel2000}. Nonetheless, even if the data are sparse the observation seems to hold for Pu as well. Furthermore, in those nuclei with a minimum $3^-_1$ energy, the energy of the 0$^+_2$ state is found at a maximum [Figs.~\ref{fig:actinidessystematics} {\bf (a)} and \ref{fig:actinidessystematics} {\bf (b)}]. Therefore, it might be possible, that strong octupole correlations in the structure of the 0$^+_2$ states are indeed correlated with close-lying 0$^+_2$ and 0$^+_3$ states in the actinides. The proposed double-octupole structure of the 0$^+_2$ state in ${}^{240}$Pu~\cite{Wang09, Jolo13} allows to study this hypothesis based on the new experimental data as a theoretical case study.

A suited model to investigate the interplay of collective quadrupole and octupole excitations in the actinides is the phenomenological approach of the {\it spdf}-IBM. The model has already shown that it can reproduce the large number of observed 0$^+$ states in the actinides \cite{Wirt00, Levon09, Levon13} by incorporating the double-octupole states. It has also proven its capability to describe the low-lying collective states in the actinides \cite{Zamf01, Zamf03}. Furthermore, in Ref. \cite{Levon13} the hadronic and electromagnetic data on ${}^{228}$Th were simultaneously described for the first time, while a double-octupole structure of the $J^{\pi}$=~0$^+_2$ state in ${}^{228}$Th was predicted~\cite{Levon13}. The same Hamiltonian and electromagnetic operators of Refs.~\cite{Zamf01, Zamf03} have been used for the present study. The dipole interaction $\hat{D}_{spdf}^{\dagger} \cdot \hat{D}_{spdf}$ was added to the Hamiltonian (Eq.~(\ref{eq:spdfhamiltonian})) since the choice of parameters for the quadrupole operator $\hat{Q}_{spdf}$ does not change the number of negative-parity bosons. Therefore, no $E2$ transitions between states built of only positive- or negative-parity bosons would be possible, in contradiction with experiment~\cite{Zamf01, Zamf03}.

\begin{equation}
\begin{split}
H~=~\epsilon_{d}\hat{n}_{d} + \epsilon_{p}\hat{n}_{p} + \epsilon_{f}\hat{n}_{f} - \kappa \hat{Q}_{spdf} \cdot \hat{Q}_{spdf} 
\\
+ \alpha \hat{D}_{spdf}^{\dagger} \cdot \hat{D}_{spdf} + H.c.
\label{eq:spdfhamiltonian}
\end{split}
\end{equation}

The parameters of the Hamiltonian are: $\epsilon_{\mathrm{d}}$=0.25, $\epsilon_{\mathrm{p}}$=1.5, $\epsilon_{\mathrm{f}}$=0.79~MeV, while the quadrupole- and dipole-coupling strenghts $\kappa$ and $\alpha$ are set to 0.013 and 0.0006~MeV, respectively. The value of the latter indicates that $\alpha$ has only a modest impact on general properties of the spectrum, but admixes a small contribution of negative-parity bosons to every state. The number of negative-parity bosons ($<\hat{n}_{\mathrm{p}}> + <\hat{n}_{\mathrm{f}}>$) in the ground states of ${}^{242,240}$Pu is smaller than 0.02, and no static octupole deformation is observed in agreement with experiment~\cite{Wied99}.

\begin{figure}[ht]
\centering
\includegraphics[width=0.72\linewidth]{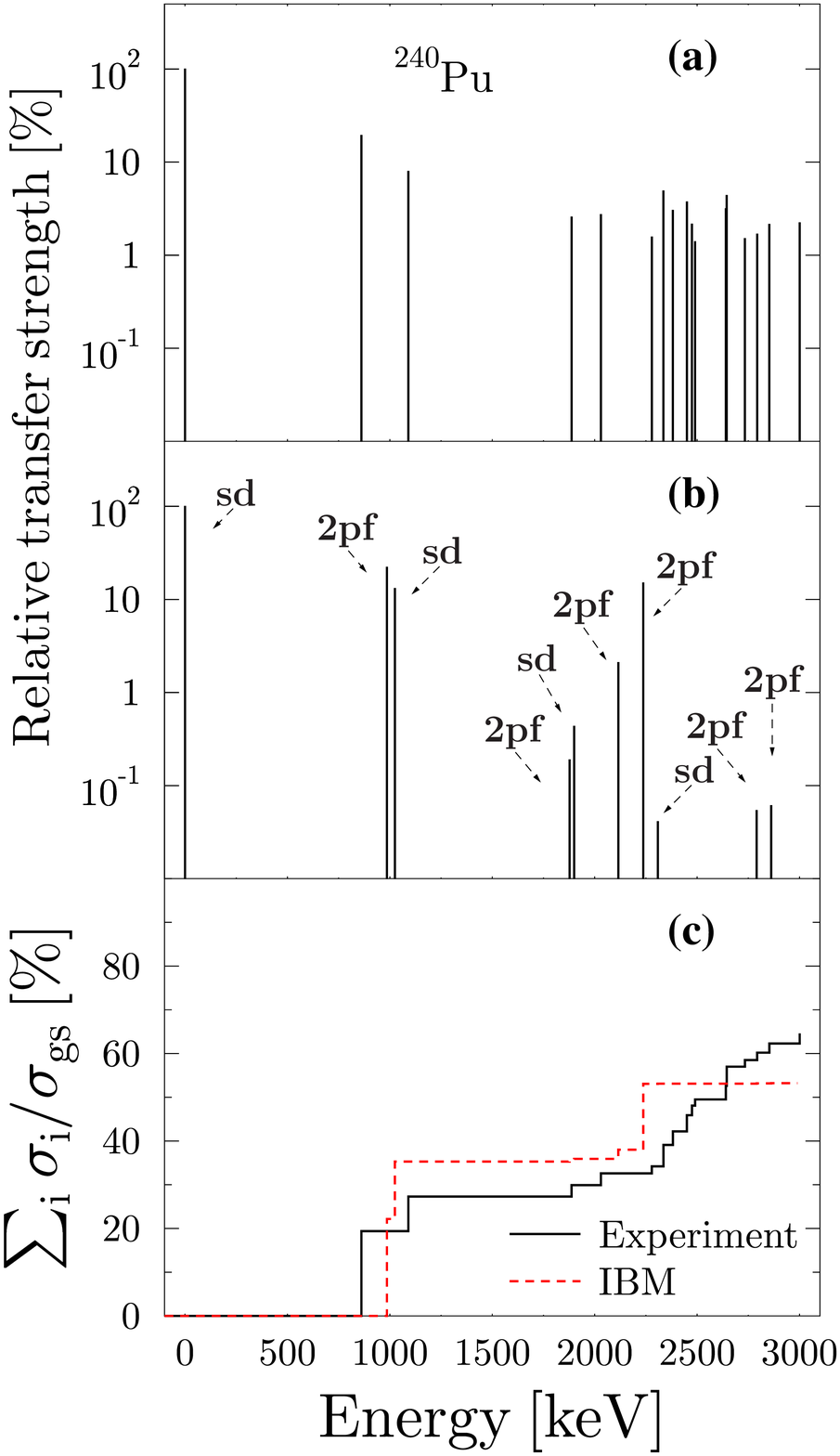}
\caption{\label{fig:0+ibmexp}(Color online) {\bf (a)} Experimental and {\bf (b)} {\it spdf}-IBM distribution of the relative transfer strength for (firmly assigned) 0$^+$ states in ${}^{240}$Pu in logarithmic scale. Furthermore, the boson structure predicted by the IBM is added to Fig. \ref{fig:0+ibmexp} {\bf (b)}. While sd highlights a quadrupole structure of the excited 0$^+$ state, 2pf indicates the presence of two negative-parity bosons in the wave function. Most of the double-dipole/octupole states have a predominant double-octupole structure. In panel {\bf (c)} the running sum of the transfer strength as a function of energy is shown.}
\end{figure}

The essential observables of experimental {\it (p,t)} studies are relative transfer strengths. In the framework of the IBM, the two-neutron transfer proceeds via the removal of an $s$-boson, and the transfer operator can be defined as follows:

\begin{equation}
\centering
\begin{split}
\hat{P}_{-,\nu}^{(0)}~=~\alpha_{\nu} \left( \Omega_{\nu} - N_{\nu} - \frac{N_{\nu}}{N} \hat{n}_{\mathrm{d}} \right)^{1/2} \left(\frac{N_{\nu}+1}{N+1}\right)^{1/2}\hat{s} 
\\
+~\left(\alpha_{\mathrm{p }} \hat{n}_{\mathrm{p}} + \alpha_{\mathrm{f}} \hat{n}_{\mathrm{f}} \right) \hat{s}~,
\end{split}
\label{ibmtransfer}
\end{equation}

where $\mathrm{\Omega_{\nu}}$ is the effective neutron pair degeneracy of the neutron shell [$\mathrm{\Omega_{\nu}}$ = $\frac{1}{2}$(164-126) = 19], $N_{\nu}$ is the number of neutron pairs (9), $N$ the total number of bosons (15), and $\alpha_\mathrm{p}$, $\alpha_\mathrm{f}$ and $\alpha_{\mathrm{\nu}}$ are constant parameters. 164 was proposed to be a neutron sub-shell closure and adopted in the calculations of Refs. \cite{Zamf01, Zamf03}. While the first term in Eq.~(\ref{ibmtransfer}) was already introduced by Arima and Iachello~\cite{Arim77}, the second term of the transfer operator had to be added to the present study as well as to the calculations presented in Ref.~\cite{Levon13} to describe the transfer intensities in the actinides. Its interpretation is not fully understood, and thus a systematic study of all terms of the transfer operator would be instructive. For now, it is taken as phenomenologically necessary to describe the experimental data. Certainly, the second term will amplify overlapping parts of negative-parity bosons in the transfer process, and therefore accounts for possible negative-parity bosons in the structure of excited 0$^+$ states. The three parameters have been set to $\alpha_{\mathrm{p}}$=~$-$1.2, $\alpha_{\mathrm{f}}$=~0, and $\alpha_{\nu}$=~0.2~mb/sr. Finally, the IBM calculations are compared to the experimental data in Fig.~\ref{fig:0+ibmexp}.

In Figs. \ref{fig:0+ibmexp}~{\bf (a)} and {\bf (b)} the experimental and theoretical distributions of 0$^+$ states are compared. While 17 states have been populated in the experiment, only 10 out of 14 0$^+$ states receive a non-vanishing transfer strength in the IBM calculations up to 3 MeV. Nonetheless, the agreement between experiment and IBM at least for the first two excited 0$^+$ states is quite good. While experimentally relative transfer strengths of 19.39(2)$\%$ and 7.96(1)$\%$ are found, the IBM predicts 22.17$\%$ and 13.12$\%$ with respect to the ground-state strength. Furthermore, the observation of two groups of excited 0$^+$ states (E$_{\mathrm{x}}<$~1.5~MeV and E$_{\mathrm{x}}>$~1.5~MeV) is predicted correctly, and their location is reproduced fairly well. Additionally, the experimental and IBM running sum of the transfer strength are found in good agreement, even though a higher fragmentation of the transfer strength around 2.25 MeV is experimentally observed [Fig. \ref{fig:0+ibmexp} {\bf (c)}]. A very good agreement is not expected, as the present IBM calculations are lacking quasiparticle excitations~\cite{Solov1986}, which are also expected to contribute to the experimental 0$^+$ distribution~\cite{Solov1997, Iudi05}. However, at energies below the 2QP energy, the systematic {\it (p,t)} study in the actinides has shown that the experimental observation of close- and low-lying 0$^+$ states in ${}^{228}$Th is not unique. Instead, two close- and low-lying 0$^+$ states are also observed in ${}^{240}$Pu and other actinides (see Fig.~\ref{fig:actinidessystematics}), therefore establishing this experimental signature and emphasizing the states' collective nature. Figure~\ref{fig:0+ibmexp}~{\bf (b)} indicates that different structures are predicted for the $J^{\pi}$=~0$^+_2$ and $J^{\pi}$=~0$^+_3$ states in ${}^{240}$Pu in agreement with experiment~\cite{Hoog96}. The IBM predicts a quadrupole structure and almost no negative-parity bosons in the wave function of the $J^{\pi}$=~0$^+_3$ state, and a double-octupole phonon structure of the $J^{\pi}$=~0$^+_2$ state as proposed in Refs. \cite{Wang09, Jolo13}. However, similar IBM calculations in Ref.~\cite{Wirt04} suggest that the large experimental 0$^+_2$ transfer strengths do not {\it per se} allow for an identification of double-octupole states, because quadrupole structures are predicted in ${}^{230}$Th and ${}^{232}$U. The observation of large relative {\it (p,t)} cross sections might therefore also point to large admixtures of pairing-correlated quasiparticles in the wave function~\cite{Iudi05, Garr01}. It is necessary to compare additional observables with the model predictions in order to test the signature of close-lying 0$^+_2$ and 0$^+_3$ states as a possible manifestation of octupole correlations in the 0$^+_2$ state of ${}^{240}$Pu and of those in other actinides. 

\begin{table}[H] 
\centering
\caption{\label{tab:be1be2}The experimental $B(E1)/B(E2)$ ratios (R$_{\mathrm{exp}}$) are compared to the predictions of the {\it spdf}-IBM (R$_{\mathrm{IBM}}$) for the low-spin members of the $K^{\pi}$=~0$^+_2$ rotational band in ${}^{240}$Pu. The energies and relative intensities are adopted from Ref.~\cite{Sing08}. No $E1$ transitions have been observed for other $K^{\pi}$=~0$^+$ bands up to now~\cite{Sing08}. The electromagnetic operators of the {\it spdf}-IBM and effective charges $e_1$ and $e_2$ in ${}^{240}$Pu are from Ref.~\cite{Zamf03}.}
\begin{ruledtabular}
\begin{tabular}{cccccc}
$J^{\pi}_{\mathrm{i}}$ & $J^{\pi}_{\mathrm{f},E1}$ & $J^{\pi}_{\mathrm{f},E2}$ & Level Energy & R$_{\mathrm{exp}}$ & R$_{\mathrm{IBM}}$\\
& & & [keV] & [10$^{-6}$~fm$^{-2}$] & [10$^{-6}$~fm$^{-2}$] \\
\hline
0$^+$ & 1$^-_1$ & 2$^+_1$ & 860.71(7) & 13.7(3) & 10.4\\
2$^+$ & 1$^-_1$ & 0$^+_1$ & 900.32(4) & 99(15) & 39.0\\
2$^+$ & 1$^-_1$ & 2$^+_1$ &  & 26(2) & 23.9\\
2$^+$ & 1$^-_1$ & 4$^+_1$ &  & 5.9(3) & 10.4\\
2$^+$ & 3$^-_1$ & 0$^+_1$ &  & 149(22) & 63.1\\
2$^+$ & 3$^-_1$ & 2$^+_1$ &  & 39(2) & 38.7\\
2$^+$ & 3$^-_1$ & 4$^+_1$ &  & 8.9(5) & 16.9\\
4$^+$ & 3$^-_1$ & 6$^+_1$ & 992.4(5) & 4.4(11) & 22.3\\
4$^+$ & 5$^-_1$ & 6$^+_1$ & & 4.7(13) & 30.6
\end{tabular}
\end{ruledtabular}
\end{table}

To further understand the structure of the 0$^+_2$ states in the actinides in detail, it is important to investigate their electromagnetic decay behavior. For ${}^{240}$Pu additional insight into the possible octupole nature of the $J^{\pi}$=~0$^+_2$ state and its rotational band members can be gained from their $\gamma$-decay branching. Especially interesting are the $B(E1)/B(E2)$ ratios of Table~\ref{tab:be1be2}, which are a direct measure for the branching to the one-octupole phonon $K^{\pi}$=~0$^-$ rotational band and to the ground-state band, respectively. Notice that these ratios are different from the $B(E1)/B(E2)$ ratios discussed in Refs.~\cite{Wied99, Wang09, Jolo11, Jolo12, Jolo13}, which were a measure for intra- and interband decays. However, since there is confidence in the double-octupole nature of the $K^{\pi}$=~0$^+_2$ rotational band~\cite{Wang09, Jolo13}, it is instructive to check the simple phonon picture by means of allowed $\Delta_{\mathrm{phonon}}$=~1 decays, i.e. the destruction of an octupole phonon. In the cases where a branching has been observed, enhanced ratios have been measured and are found in agreement with the model predictions. No $E1$ transitions have been observed for any other of the previously known $K^{\pi}$=~0$^+$ bands in ${}^{240}$Pu~\cite{Sing08}. Additionally, $E0$ transitions could be used to characterize excited 0$^+$ states in atomic nuclei~\cite{Garr01, VonBrentano04, Iudi05}. To calculate $E0$ transitions only the $d$-boson contribution of the operator was considered and the corresponding parameter was set to 0.053~\cite{Iach87, VonBrentano04}. The present results have been normalized to the measured $X(E0/E2)$ ratio of 0.048(16) of the $J^{\pi}$= 0$^+_2$ state~\cite{Sing08}. They support the observation of a less enhanced $0^+_2 \rightarrow 0^+_{\mathrm{1}}$ $E0$ transition compared to the $0^+_3 \rightarrow 0^+_{\mathrm{1}}$ $E0$ transition in ${}^{240}$Pu~\cite{Hoog96}. A suppression by a factor of 5 was estimated in Ref.~\cite{Hoog96}. The present IBM calculation predicts a factor of 13. Therefore, the double-octupole structure and the necessary annihilation of two octupole phonons at once might yield a possible explanation of the experimentally observed suppression of the $0^+_2 \rightarrow 0^+_{\mathrm{1}}$ $E0$ transition.

In summary, a high-resolution ${}^{242}$Pu{\it (p,t)}${}^{240}$Pu experiment has been performed to study 0$^+$ states up to $E_{\mathrm{x}}$=~3.1~MeV. It has been shown that the {\it spdf}-IBM can account for both hadronic and electromagnetic observables in ${}^{240}$Pu, while two excited 0$^+$ states of different structure are predicted below the 2QP energy. Consequently, the proposed octupole correlations in the structure of the $K^{\pi}$=~0$^+_2$ band in ${}^{240}$Pu~\cite{Wang09, Jolo13} are supported. The present study indicates that the structure of some $J^{\pi}$=~0$^+_2$ states in the actinides might be more complex than the usually discussed pairing isomers or pairing vibrations. Especially octupole correlations in their structure might give a possible explanation for the experimentally observed fast $E1$ $\gamma$ decays to the one-octupole phonon states~\cite{Jolo13}. Several actinide nuclei are also accessible by the {\it (p,t)} and {\it (t,p)} reactions~\cite{Maher72, Cast72, Fried74} and could be remeasured in state-of-the-art experiments. New data might help to discriminate between pairing isomers, which could mix with quadrupole one-phonon excitations, and possible double-octupole excitations. The present results on the latter could contradict earlier conclusions about their general existence~\cite{Solov1981}. It is therefore necessary to clarify whether dominant contributions or small admixtures~\cite{Weber98} in the wave function are needed or if other structures are able to explain the experimental observables presented here. Promising candidates to search for manifestations of octupole correlations in the 0$^+_2$ states are ${}^{224,226}$Ra~\cite{Gaff13}, ${}^{232}$Th~\cite{Jolo13}, ${}^{238}$U~\cite{Zhu10}, and ${}^{238}$Pu~\cite{Wied99} as expected from close-lying 0$^+_2$ and 0$^+_3$ states presented in Fig.~\ref{fig:actinidessystematics}. In the future, the study of exotic nuclei in inverse kinematics might also be possible at the CERN On-Line Isotope Mass Seperator (ISOLDE) facility~\cite{Gaff13, Wimm10}. Furthermore, it appears that octupole correlations might also be observed in some 0$^+_2$ states of the rare-earth nuclei~\cite{Bvum13}.

We gratefully acknowledge the help of the accelerator staff at MLL in Munich. Furthermore, we thank P.~von Brentano, R.F.~Casten, V.~Derya, J.~Jolie, R.V.~Jolos, L.~Netterdon, S.G.~Pickstone, and N.~Warr for helpful discussions. This work is supported in part by the Deutsche Forschungsgemeinschaft under Contract ZI 510/4-2 and the Romanian UEFISCDI Project No. PN-II-ID-PCE-2011-3-0140.

\bibliography{240Pu_prcr}

\end{document}